\documentclass[11pt,a4paper,twoside]{article}
\usepackage{graphicx}
\usepackage{color}
\usepackage{epstopdf}
\usepackage{lscape}
\usepackage{amsmath}
\usepackage{hyperref}
\usepackage{setspace}
\begin{document}
\baselineskip=0.30in
\begin{center}
{\Large \bf Solutions of the Dirac equation with the shifted Deng-Fan potential including Yukawa-like tensor interaction} 
\end{center}

\begin{center}
{\bf  W. A. Yahya \footnote{\scriptsize E-mail:~ wazzy4real@yahoo.com}}\\
\footnotesize{\it
 Department of Physics and Material Science,\\
Kwara State University, P. M. B. 1530, Ilorin, Nigeria.}\\

{\bf B. J. Falaye \footnote{\scriptsize E-mail:~ fbjames11@physicist.net}}\\
\footnotesize{\it
Theoretical Physics Section, Department of Physics,\\
 University of Ilorin,  P. M. B. 1515, Ilorin, Nigeria.}\\
{\bf O. J. Oluwadare \footnote{\scriptsize E-mail:~ jostimint@yahoo.com}} \\
\footnotesize{\it
Department of Physics, Federal University Oye-Ekiti,\\
P. M. B. 373, Ekiti State, Nigeria.}\\

{\bf K. J. Oyewumi \footnote{\scriptsize E-mail:~ kjoyewumi66@unilorin.edu.ng} }\\
\footnotesize{\it Theoretical Physics Section, Department of Physics,\\
 University of Ilorin,  P. M. B. 1515, Ilorin, Nigeria.}\\
\end{center}

\begin{abstract}
\noindent 
By using the Nikiforov-Uvarov method, we give the approximate analytical solutions of the Dirac equation with the shifted Deng-Fan potential including the Yukawa-like tensor interaction under the spin and pseudospin symmetry conditions. After using an improved approximation scheme, we solved the resulting schr\"{o}dinger-like equation analytically. Numerical results of the energy eigenvalues are also obtained, as expected, the tensor interaction removes degeneracies between spin and pseudospin doublets.
\end{abstract}


\section{\bf Introduction}
Pseudospin was introduced in the late 60's \cite{ArE69, HeA69} and it has been used to explain features of deformed nuclei \cite{BoE82} and superdeformation \cite{DuE87}, and to establish an effective shell-model coupling scheme \cite{TrE95}. Spin and pseudospin symmetry now play an important role in Hadron and nuclear spectroscopy \cite{Gin051, GiL98, Mao03, AlE05, FuE98, NaE90}.

In recent years, a lot of attention have been paid to Dirac equation with spin and pseudospin symmetry and their applications. For example, Zhou et al \cite{ZhE03} used the relativistic mean field theory to investigate single anti-nucleon spectra. The spin symmetry in antinucleon spectra of a nucleus have been tested by investigating the relations between the Dirac wave functions of the spin doublets and examining these relations in realistic nuclei within the relativistic mean-field model \cite{HeE06}. From the Dirac equation, the mechanism behind the pseudospin symmetry was studied by Meng et al \cite{MeE98, MeE99} and the pseudospin symmetry was shown to be connected with the competition between the centrifugal barrier  and the pseudospin orbital potential, which is mainly decided by the derivative of the difference between the scalar and vector potentials. 

Very recently, Song et al. \cite{SoE11} studied the effects of tensor coupling on the spin symmetry of $\bar{\Lambda}$ spectra in $\bar{\Lambda}$-nucleus systems with the relativistic mean-field theory. Shortly thereafter, Lu et al. \cite{LuE12} show that the conservation and the breaking of the pseudospin symmetry in resonant states and bound states share some similar properties. Furthermore in 2013, By examining the zeros of Jost functions corresponding to the small components of Dirac wave functions and phase shifts of continuum states, Lu et al. \cite{LuE12} also showed that the pseudospin symmetry in single particle resonant states in nuclei is conserved when the attractive scalar and repulsive vector potentials have the same magnitude but opposite sign \cite{LuE13}. Li et al. \cite{LiE13} proved that the spin-orbit interactions always play a role in favor of the pseudospin symmetry, and whether the pseudospin symmetry is improved or destroyed by the dynamical term relating the shape of the potential as well as the quantum numbers of the state.

In recent years, some authors have investigated the spin symmetry and Pseudospin symmetry under the Dirac equation for some typical potentials such as the Harmonic oscillator potential \cite{LiE04, Gin052, GoE05, CaE06, SeN10, LiE11, Akc07, AkT09, Akc09}, Coulomb potential \cite{HamE101, ZaE11}, Woods-Saxon potential \cite{GoS05, AyS10}, Morse potential \cite{DoS03, FiR00, QiE07, BaB07, Ber06, HamE121}, Eckart potential \cite{JiE06, ZhE08}, ring-shaped non-spherical harmonic oscillator \cite{GoE06}, P\"{o}schl-Teller potential \cite{JiE071, WeD11, HasE121, JiE09, FaI13}, three parameter potential function as a diatomic molecule model \cite{JiE072}, Yukawa potential \cite{SeH10, HamE122, AyS11, MaE121, IkF13}, pseudoharmonic potential \cite{AyS101}, Davidson potential \cite{SeH09}, Mie-type potential \cite{HamE102}, Deng-Fan potential \cite{MaE122}, hyperbolic potential \cite{HasE11}, Tietz potential \cite{HasE122} and Rosen-Morse potential \cite{OyA10, Oye121, Oye122}.  

Different methods have been used such as Supersymmetric quantum mechanics, Nikiforov-Uvarov method, Path integral, series expansion, Ansatz techniques, etc. In this paper, we have used the elegant Nikiforov-Uvarov method to obtain the improved approximate analytical solutions of the Dirac equation under the  Deng-Fan Scalar and Vector potentials and a coulomb-like tensor. The improved approximate analytical solutions can be obtained by using the improved new approximation scheme (Pekeris-type approximation scheme, Pekeris (1934) \cite{Pek34}) to deal with the centrifugal (pseudo-centrifugal) term near the minimum point $r = r_{e}$ \cite{Ikh11}. In addition, the corresponding lower and upper component spinors are obtained. Lalazissis et al. \cite{LaE98} investigated the pseudospin approximation in exotic nuclei in $Zr$ and $Sn$ isotopes from the proton drip line to the neutron drip line based on the relativistic continuum Hartree-Bogoliubov theory. Nuclear halo structure and conservation of relativistic symmetry have been studied within the framework of the relativistic Hartree-Fock-Bogoliubov theory by Long et al. \cite{LoE10}.

Deng - Fan molecular potential, which is an exponential-type potential, is a simple modified Morse potential called generalized Morse potential which was proposed by Deng and Fan in (1957) \cite{DeF57}, in an attempt to find a more suitable diatomic potential to describe the vibrational spectrum \cite{Don07}. This potential has been widely studied by some researchers in various applications \cite{Ikh11, DeF57, Don07, RoE03,   MeE98b, ZhE09, RoS09, Don11, OlE121,  OlE122, OyE131}. This potential model can be used to describe the motion of the nucleons in the mean field produced by the interactions between nuclei \cite{Ikh11}.  Maghsoodi et al. (2012) obtained the spectrum of Dirac equation under the Deng-Fan potential with a tensor interaction, using supersymmetric quantum mechanics \cite{MaE122}. 

Very recently, Hamzavi et al. (2013) \cite{HamE13} suggested a modified form of the Deng-Fan potential, called the shifted Deng-Fan potential. The modification is the Deng-Fan potential shifted by dissociation energy $\bar {\alpha}$ \cite{HamE13}. In their work, it was demonstrated that shifted Deng-Fan potential and the Morse potential are very close to each other for large values of $r$ in the regions $ r \approx r_{e}$ and $r>r_{e}$, but different at $r \approx 0$ \cite{HamE13}. Also, the thermodynamic properties and the approximate solutions of the Schr\"{o}dinger Equation with this potential model have been studied by Oyewumi et al. (2013) \cite{OyE132}. 

In order to obtain theoretical results that are comparable with the experimental results, a tensor interaction is always been introduced in the Dirac equation with the Pseudospin and spin symmetry limits to remove the degeneracy in the pseudospin and spin doublets \cite{Akc07, AkT09, Akc09, ItE69, MoS89, KuE91,  HasE123}. Tensor couplings or interactions have been used successfully in the studies of nuclear properties and applications \cite{Mao03, AlE05, FuE98, CaE06,  Akc07, AkT09, Akc09, HamE101, ZaE11,   HamE121, HasE121,  HamE122, AyS11, MaE121, IkF13, AyS101,  HamE102, MaE122, HasE11, HasE122, Oye122, ItE69,  MoS89, KuE91, HasE123, IkS10,  LiE041, LiE042, LiE043, ZaE10,  EsH12, Esh111, EsM111, EsM112,  AyS104, Ayd09, AyE13}.

In most of these investigations, Coulomb-like tensor interaction has been used, except in a few instances where linear plus Coulomb tensor interaction has been used \cite{HasE121, ZaE10}. Recently, another form of tensor interaction (Yukawa potential as a tensor interaction) has been introduced by  Hassanabadi et al. (2012) and Aydo$\check{g}$du et al. (2013) \cite{HasE123, AyE13}.

It is therefore the purpose of the present study to investigate the Dirac equation with the shifted Deng-Fan potential coupled with the Yukawa tensor potential (instead of the usual Coulomb-like tensor) under the pseudospin and spin symmetry limits. The approximate scheme (the Pekeris-type approximation scheme) has been used to deal with the (pseudo-) centrifugal term, this problem has been solved via the Nikiforov-Uvarov method.

The scheme of our presentations are as follows: In Section $2$, we give a brief review of the Dirac equation with Yukawa tensor interaction. Section $3$ contains the approximate solution to the problem.  The Conclusions and  some numerical results for the energy spectrum are given in section $4$.

\section{Dirac bound state solutions including Yukawa-like Tensor}
The time independent Dirac equation with the scalar potential $S(r)$, vector potential $V(r)$ and a tensor potential $U(r)$ is
written for a spin-half particle in the natural units $\hbar=c=1$ as \cite{Gin051, Gin052, IkF13, MaE122,  Oye121, Oye122, Gre00}:
\begin{equation}
\left[\alpha.\textbf{p}+\beta(M+S(r))-i\beta\alpha.\widehat{r}U(r)\right]\Psi_{nk}(\textbf{r})=[E-V(r)]\Psi_{n\kappa}(\textbf{r})
\label{wr1}
\end{equation}
where E and M are the relativistic energy of the system and the rest mass of the spin-half particle, respectively. $\textbf{p}=-i\nabla$ is the three dimensional momentum operator. $\alpha$ and $\beta$ are
$4\times4$ Dirac matrices, given as:
\begin{equation}
\alpha=\left(
  \begin{array}{cc}
    0 & \sigma_{i} \\
    \sigma_{i} & 0 \\
  \end{array}
\right),\;\;    \beta=\left(
  \begin{array}{cc}
    I & 0 \\
    0 & -I \\
  \end{array}
\right)
\label{wr2},
\end{equation}
where $I$ is the $2\times2$ identity matrix and $\sigma_{i}(i=1,2,3)$ are the three $2\times2$ Pauli spin matrices
\begin{equation}
\sigma_{1}=\left(
             \begin{array}{cc}
               0 & 1 \\
               1 & 0 \\
             \end{array}
           \right),\;\; \sigma_{2}=\left(
                                     \begin{array}{cc}
                                       0 & -i \\
                                       i & 0 \\
                                     \end{array}
                                   \right),\;\; \sigma_{3}=\left(
                                                             \begin{array}{cc}
                                                               1 & 0 \\
                                                               0 & -1 \\
                                                             \end{array}
                                                           \right).
\label{wr3}
\end{equation}
For a particle in a spherical (central) field, the total angular momentum operator $\textbf{J}$ and the spin-orbit matrix operator $\hat{\textbf{K}}=-\beta(\sigma.\textbf{L}+\textbf{I})$ commute with the Dirac Hamiltonian, where $\textbf{L}$ is the orbital angular
momentum operator. For a given total angular momentum $j$, the eigenvalues of $\hat{\textbf{K}}$ are $\kappa=-(j+\frac{1}{2})$ for
aligned spin $(s_{\frac{1}{2}}, p_{\frac{3}{2}}, etc.)$ and $\kappa=j+\frac{1}{2}$ for unaligned spin $(p_{\frac{1}{2}}, d_{\frac{3}{2}}, etc)$. The spinor wave functions can be classified according to the radial quantum number, n and the spin-orbit quantum number $\kappa$ and can be written using the Pauli-Dirac representation \cite{Gin051, GiL98,  Gin052}
\begin{equation}
\Psi_{n\kappa}(\textbf{r})= \frac{1}{r} \left[\begin{matrix} 
F_{n\kappa}(\textbf{r})Y_{jm}^{\ell}(\theta, \vartheta)\\
iG_{n\kappa}(\textbf{r})Y_{jm}^{\tilde{\ell}}(\theta, \vartheta)
\end{matrix}\right],~~ 
\label{wr4}
\end{equation}
where $F_{n\kappa}(\textbf{r})$ and $G_{n\kappa}(\textbf{r})$ are the radial wave functions of the upper and lower spinor components,
respectively. $Y_{jm}^{\ell}(\theta, \vartheta)$ and
$Y_{jm}^{\tilde{\ell}}(\theta, \vartheta)$ are the spherical
harmonics functions coupled to the total angular momentum j and its
projection m on the z-axis. The orbital and pseudo-orbital angular
momentum quantum numbers for spin symmetry $\ell$ and pseudo-spin
symmetry $\tilde{\ell}$ refer to the upper and lower components,
respectively.

Substituting equation (\ref{wr4}) into equation (\ref{wr1}), we
obtain the two coupled first-order differential equation
\begin{equation}
\left(\frac{d}{dr} + \frac{\kappa}{r} - U(r)\right)F_{_{n\kappa}}(r)
= \left( M + E_{n\kappa} - \Delta(r)\right)G_{n\kappa}(r),
\label{wr5}
\end{equation}
\begin{equation}
\left(\frac{d}{dr} - \frac{\kappa}{r}+U(r)\right)G_{_{n\kappa}}(r) =
\left( M - E_{n\kappa} + \Sigma(r)\right)F_{n\kappa}(r),
\label{wr6}
\end{equation}
where $\Delta(r)=V(r)-S(r)$ and $\Sigma(r)=V(r)+S(r)$ are the
difference and the sum potentials, respectively. By eliminating
$G_{n\kappa}(r)$ in equation (\ref{wr5}) and $F_{n\kappa}(r)$ in
equation (\ref{wr6}), we obtain two second-order non-linear
differential equations for the upper and lower radial spinor
components as, respectively
\begin{eqnarray}
&\displaystyle \left[\frac{d^{2}}{dr^{2}} -
\frac{\kappa(\kappa+1)}{r^{2}}+
\frac{2\kappa}{r}U(r)-\frac{dU(r)}{dr} - U^{2}(r) +
\frac{\frac{d\Delta(r)}{dr}}{M+E_{n\kappa}-\Delta(r)}\left(\frac{d}{dr}+\frac{\kappa}{r}-U(r)\right)\right]F_{n\kappa}(r)
\nonumber \\
&
\displaystyle=\left[(M+E_{n\kappa}-\Delta(r))(M-E_{n\kappa}+\Sigma(r))\right]F_{n\kappa}(r),
\label{wr7}
\end{eqnarray}
\begin{eqnarray}
&\displaystyle \left[\frac{d^{2}}{dr^{2}} -
\frac{\kappa(\kappa-1)}{r^{2}}+
\frac{2\kappa}{r}U(r)+\frac{dU(r)}{dr} - U^{2}(r) +
\frac{\frac{d\Sigma(r)}{dr}}{M-E_{n\kappa}+\Sigma(r)}\left(\frac{d}{dr}-\frac{\kappa}{r}+U(r)\right)\right]G_{n\kappa}(r)
\nonumber \\
&
\displaystyle=\left[(M+E_{n\kappa}-\Delta(r))(M-E_{n\kappa}+\Sigma(r))\right]G_{n\kappa}(r),
\label{wr8}
\end{eqnarray}
where $\kappa(\kappa-1)=\tilde{\ell}(\tilde{\ell}+1)$ and $\kappa(\kappa+1)=\ell(\ell+1).$

\subsection{Spin Symmetry Limit}
The spin symmetry occurs when $\frac{d\Delta (r)}{dr}=0$ or $\Delta (r)=C_{s}=$ constant \cite{Gin051, GiL98, Gin052,  ZhE08, MaE122, Ikh11}. Here we are taking $\Sigma(r)$ as the shifted Deng-Fan potential \cite{HamE13}, i.e. 
\begin{equation}
\Sigma(r)=Db\left[\frac{b}{\left(e^{ar}-1\right)^2}-\frac{2}{e^{ar}-1}\right],\ \ \ \ \ b=e^{ar_{e}}-1
\label{wr9}
\end{equation}
where $r \in(0,\infty)$ and the three positive parameters $D$, $r_e$ and $a$ denote the dissociation energy,
the equilibrium inter-nuclear distance, and the range of the potential well, respectively. The Yukawa tensor potential  is introduced instead of the usual Coulomb-like tensor interaction as
\begin{equation}
    U(r)=-\frac{A}{r}e^{-ar}
		\label{wr10}
\end{equation}
where $A=\gamma Z$, $\gamma=(137.037)^{-1}$ is the fine-structure constant, $Z$ is the atomic number. This potential is often used to compute bound-state normalizations and energy levels of neutral atoms \cite{IkF13, McE76}. Under this symmetry, equation (\ref{wr7}) can easily be transformed to 
\begin{eqnarray}
\left[\frac{d^{2}}{dr^{2}}-\frac{\kappa(\kappa+1)+(2\kappa+1)Ae^{-ar}+A^2e^{-2ar}}{r^{2}}-\frac{Aae^{-ar}}{r}\right]F_{s,n\kappa}(r)\nonumber\\
\left[(M+E_{n\kappa}-C_s)\left(M-E_{n\kappa}+Db\left(\frac{b}{\left(e^{ar}-1\right)^2}-\frac{2}{e^{ar}-1}\right)\right)\right]=0
\label{wr11}
\end{eqnarray}
where $\kappa =\ell $ and $\kappa =-\ell -1$ for $\kappa <0$ and $\kappa >0$%
, respectively

\subsection{Pseudospin Symmetry Limit}

The pseudospin symmetry occurs when $\frac{d[V(r)+S(r)]}{dr}=%
\frac{d\Sigma (r)}{dr}=0$ or $\Sigma (r)=C_{ps}=$constant. Here we are taking $\Delta (r)$ as the shifted Deng-Fan potential and the tensor as the Yukawa-like potential, i.e. 
\begin{equation}
\Delta(r)=Db\left[\frac{b}{\left(e^{ar}-1\right)^2}-\frac{2}{e^{ar}-1}\right],\ \ \ \ \ \ \mbox{and}\ \ U(r)=-\frac{A}{r}e^{-ar},\ \ \ \ r\geq R_{c}, 
\label{wr12}
\end{equation}
By substituting equation (\ref{wr12}) into equation (\ref{wr8}), we can easily find
\begin{eqnarray}
 \left[\frac{d^{2}}{dr^{2}}-\frac{\kappa(\kappa-1)+(2\kappa+1)Ae^{-ar}+A^2e^{-2ar}}{r^{2}}-\frac{Aae^{-ar}}{r}\right]G_{ps,n\kappa}(r)\nonumber\\
\left[(M-E_{n\kappa}+C_{ps})\left(M+E_{n\kappa}-Db\left(\frac{b}{\left(e^{ar}-1\right)^2}-\frac{2}{e^{ar}-1}\right)\right)\right]=0,
\label{wr13}
\end{eqnarray}
where $\kappa =-\tilde{\ell}$ and $\kappa =\tilde{\ell}+1$ for $\kappa <0$ and $\kappa>0$, respectively.

\section{Bound state solutions of the Dirac equation with the Shifted Deng-Fan potential and the Yukawa tensor potential}

In this section, by employing the NU method, we shall approximately solve the Dirac equation with the shifted Deng-Fan potential and the Yukawa tensor potential.

\subsection{Solutions of spin symmetry limit}
In order to obtain the exact solution of equation (\ref{wr11}) for $\kappa\neq (0,-1)$, we must therefore include an approximation to deal with the centrifugal barrier. It is found that for short-range potential, the following approximation is a good approximation to the spin/pseudospin orbit coupling term
\begin{equation}
    \frac{1}{r^2}\approx\frac{a^2}{\left(1-e^{-ar}\right)^2},
		\label{wr14}
\end{equation} 
\begin{figure}[!htb]
\centering \includegraphics[height=100mm,width=180mm]{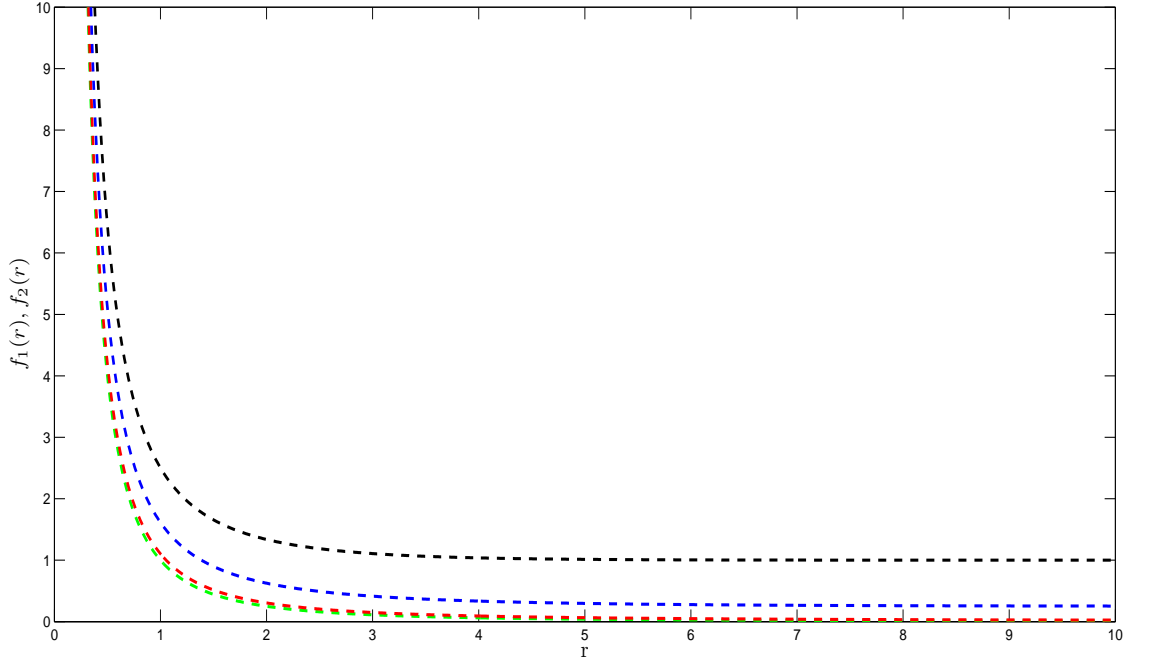}
\caption{{\protect\footnotesize The plots of expressions $f_1(r)=\frac{1}{r^2}$ (green dash line) and $f_2(r)=\frac{a^2}{\left(1-e^{-ar}\right)^2}$ as a function of the variable r are displayed with parameters $a=0.1, 0.5, 1$ correspond to red dash, blue dash and black dash lines respectively.}}
\label{fig1}
\end{figure}
 In order to show the rationality and validity of the approximation \ref{wr14}, we define the following two functions similarly to Dong \cite{Don11} and Falaye \cite{Fal12a}
\begin{equation}
f_1(r)=\frac{1}{r^2}\ \ ,\ \ \ f_2(r)=\frac{a^2}{\left(1-e^{-ar}\right)^2}\ ,\ \ \ \ a=0.1, 0.5, 1.0,
\label{APROX}
\end{equation}
and show the plots of expressions $f_1(r)$ and $f_2(r)$ as a function of the variable r with parameter $a=0.1, 0.5, 1.0$ in figure \ref{fig1}. It is shown that when $a$ is small (short potential range), equation (\ref{wr14}) is a good approximation to $\frac{1}{r^2}$. However, the difference between them will appear for large $a$ (long-range potential). This implies that the equation (\ref{wr14}) is not a good approximation to the centrifugal term when the parameter $a$ becomes large. Now, with the aid of equation (\ref{wr14}) and by introducing the following new dimensionless parameter, $z(r) = e^{-ar}$ $\in[0, 1]$, which maintains the finiteness of the transformed wave functions on the boundary conditions, equation (\ref{wr11}) is easily transformed into the following Schr\"{o}dinger-like equation satisfying $F_{s,n\kappa}(r)$
\begin{equation}
\frac{d^{2}F_{s,n\kappa}}{dz^{2}}+\frac{1}{z}\frac{dF_{s,n\kappa}}{dz}-\left\{\frac{\left(A^2-A+\xi(b+2)+\zeta\right)z^{2}+\left(2A(\kappa+1)-2\xi-2\zeta\right)z+\kappa(\kappa+1)+\zeta}{z^{2}(1-z)^{2}}\right\}F_{s,n\kappa}(z)=0,
\label{wr15}
\end{equation}
where we have have introduced the following parameters
\begin{equation}
\xi=\frac{1}{a^2}\left(M+E_{n\kappa}-C_{s}\right)Db,\ \ \ \  \zeta=\frac{1}{a^2}\left(M+E_{n\kappa}-C_s\right)\left(M-E_{n\kappa} \right).
		\label{wr16}
\end{equation}
Equation (\ref{wr15}) can now be solved by using the Nikiforov-Uvarov method, and thus, the required polynomials given in appendix A take the following analytic forms:
\begin{subequations}
\begin{equation}
    \pi(z)=-\frac{z}{2}-\left[\left[\sqrt{\left(A+\kappa+\frac{1}{2}\right)^2+\xi b}+\sqrt{\kappa(\kappa+1)+\zeta}\right]z-\sqrt{\kappa(\kappa+1)+\zeta}\right],
		\label{wr17a}
\end{equation}
\begin{equation}
		 k=2\xi-2(A+\kappa)(k+1)-2\sqrt{\left(\kappa(\kappa+1)+\zeta\right)\left[\left(A+\kappa+\frac{1}{2}\right)^2+\xi b\right]},
\end{equation}
\begin{equation}
		\tau(z)=1-2z-2\left[\left[\sqrt{\left(A+\kappa+\frac{1}{2}\right)^2+\xi b}+\sqrt{\kappa(\kappa+1)+\zeta}\right]z-\sqrt{\kappa(\kappa+1)+\zeta}\right],
\end{equation}
\end{subequations}
with $\tau'(z)=-2-2\left[\sqrt{\left(A+\kappa+\frac{1}{2}\right)^2+\xi b}+\sqrt{\kappa(\kappa+1)+\zeta}\right]<0$. It is straight forward to established the following relation by using (\ref{A10})
\begin{equation}
\left(n+\frac{1}{2}+\sqrt{\left(A+\kappa+\frac{1}{2}\right)^2+\xi b}+\sqrt{\kappa(\kappa+1)+\zeta}\right)^2-A^2+A-\xi(b+2)-\zeta=0
\label{wr18}
\end{equation}
By using the identities in equation (\ref{wr16}), we can find a more explicit expression for the relativistic bound-state energy spectrum as
\begin{eqnarray}
&\left(n+\frac{1}{2}+\sqrt{\left(A+\kappa+\frac{1}{2}\right)^2+\frac{Db^2}{a^2}\left(M+E_{n\kappa}-C_{s}\right)}+\sqrt{\kappa(\kappa+1)+\frac{1}{a^2}\left(M+E_{n\kappa}-C_s\right)\left(M-E_{n\kappa} \right)}\right)^2 \\ \nonumber 
& -A^2+A-\frac{Db(b+2)}{a^2}\left(M+E_{n\kappa}-C_{s}\right)-\frac{1}{a^2}\left(M+E_{n\kappa}-C_s\right)\left(M-E_{n\kappa} \right)=0.
\label{wr19}
\end{eqnarray}
In order to the obtain the corresponding wave functions, we find the explicit form of the weight function (Appendix \ref{A10}) as
\begin{equation}
    \rho(z)=z^{2\delta}\left(1-z\right)^{\eta},\ \ \ \delta=\sqrt{\kappa(\kappa+1)+\zeta}\ \ \ \eta=2\sqrt{\left(A+\kappa+\frac{1}{2}\right)^2+\frac{Db^2}{a^2}\left(M+E_{n\kappa}-C_{s}\right)}
		\label{wr20}.
\end{equation}
The above weight function gives the first part of the wave functions from (Appendix \ref{A9}) as:
\begin{equation}
y_{n\kappa}(z)\rightarrow P_{n}^{\left(2\delta,\ \eta\right)}(1-2z),
\label{wr21}
\end{equation}
and hence the second part of the wave functions can be easily found by using (Appendix \ref{A5}) as
\begin{equation}
    \phi(z)\rightarrow z^\delta\left(1-z\right)^{\frac{\eta}{2}+\frac{1}{2}}.
\label{wr22}
\end{equation}
Hence, the unnormalized upper component  of the spinor wave functions is expressed in terms of the Jacobi polynomials as (Appendix \ref{A2})
\begin{equation}
F_{s,n\kappa}(r)=N_{s,n\kappa}\ z^\delta\left(1-z\right)^{\frac{\eta}{2}+\frac{1}{2}}P_{n}^{\left(2\delta,\ \eta\right)}(1-2z)
\label{wr23}
\end{equation}
where $N_{s,n\kappa}$ is the normalization constant. 


\subsection{Solutions of pseudospin symmetry limit}
The lower component  of the spinor wave function $G_{ps,n\kappa}$ in equation (\ref{wr13}) can be obtained by following the same procedures in the previous subsection, this equation can be expressed as 
{\small\begin{equation}
\frac{d^2G_{ps,n\kappa}(z)}{dz^2}+\frac{1}{z}\frac{dG_{ps,n\kappa}(z)}{dz}-\left\{\frac{\left(A^2-A+\tilde{\xi}(b+2)+\tilde{\zeta}\right)z^2+\left(2A(\kappa+1)-2\tilde{\xi}-2\tilde{\zeta}\right)z+\kappa(\kappa-1)+\tilde{\zeta}}{z^{2}(1-z)^2}\right\}G_{ps,n\kappa}(z)=0,
\label{wr24}
\end{equation}}
where the following parameters have been introduced:
\begin{equation}
    \tilde{\xi}=\frac{1}{a^2}\left(E_{n\kappa}-M-C_{s}\right)Db,\ \ \ \  \tilde{\zeta}=\frac{1}{a^2}\left(M-E_{n\kappa}+C_s\right)\left(M+E_{n\kappa}\right).
		\label{wr25}
\end{equation}
To avoid repetition in the procedure of obtaining the solution for this case, the following parameter mapping is used \cite{Ikh10}:
\begin{equation}
F_{s, n\kappa}(r)\leftrightarrow G_{ps,n\kappa}, \ \ \kappa\rightarrow\kappa-1, \ \ V(r)\rightarrow-V(r), \ \ E_{s,n\kappa}\rightarrow -E_{ps,n\kappa}, \ \ C_s\rightarrow -C_{ps}.
\label{wr27}
\end{equation}
By applying the above transformations to equations, equations (\ref{wr19}), (\ref{wr23}) and (\ref{wr24}) lead to the following pseudospin symmetric energy equations 
\begin{eqnarray}
&&\left(n+\frac{1}{2}+\sqrt{\left(A+\kappa-\frac{1}{2}\right)^2+\frac{Db^2}{a^2}\left(E_{n\kappa}-M-C_{s}\right)}+\sqrt{\kappa(\kappa-1)+\frac{1}{a^2}\left(M-E_{n\kappa}+C_s\right)\left(M+E_{n\kappa} \right)}\right)^2\nonumber\\
&&-A^2+A-\frac{Db(b+2)}{a^2}\left(E_{n\kappa}-M-C_{s}\right)-\frac{1}{a^2}\left(M-E_{n\kappa}+C_s\right)\left(M+E_{n\kappa}\right)=0
\label{wr28}
\end{eqnarray}
and the unnormalized lower component  of the spinor wave function is obtained in terms of the Jacobi polynomials as
\begin{equation}
G_{ps,n\kappa}(r)=\tilde{N}_{s,n\kappa}\ z^{\tilde{\delta}}\left(1-z\right)^{\frac{\tilde{\eta}}{2}+\frac{1}{2}}P_{n}^{\left(2\tilde{\zeta},\ \tilde{\eta}\right)}(1-2z)
\label{wr29}
\end{equation}
where $\tilde{N}_{s,n\kappa}$ is the normalization constant.

\section{Conclusion}
We have obtained the approximate energy equation and the corresponding wave functions of the Dirac equation for shifted Deng-Fan potential coupled with a Yukawa-like tensor under the conditions of the spin and pseudospin symmetry. After using an improved approximation scheme, the resulting Schr\"{o}dinger-like equation is solved by using the elegant Nikiforov-Uvarov method. Our numerical data describe the energy splitting in detail. In table \ref{tab1}, in the case of exact spin, $r_e=0.4$ and absence of tensor interaction, i.e., $A=0$ (when the tensor potential vanishes), we noticed the degeneracy between two states in the spin symmetry doublets:
\begin{equation}n\left\{\begin{matrix}p_{3/2}\\d_{5/2}\\f_{7/2}\\g_{9/2}\end{matrix}\right.=n\left\{\begin{matrix}p_{1/2}\\d_{3/2}\\f_{5/2}\\g_{7/2}\end{matrix}\right.\nonumber.\end{equation}
Also, the system becomes less attractive when the quantum numbers increasing, i.e., the energy levels are negative. By decreasing equilibrium inter-nuclear distance by half yielded a very small increment in the energies. It is also noticed that the presence of tensor interaction, say $A=0.5$, removes the degeneracy between the states. When $C_s=0$, we noticed that the energy levels are repulsive and becoming weakly repulsive with quantum numbers increasing.

Furthermore, in table \ref{tab2}, we take $C_{ps}=0$(exact pseudospin) and $C_{ps}=-5$. For $C_{s}=0$, $r_e=0.8fm^{-1}$ and $A=0$, we noticed that the system is repulsive also the degeneracy between two states in the pseudospin symmetry doublets can be written as follows: 
\begin{equation}n\left\{\begin{matrix}s_{1/2}\\p_{3/2}\\d_{5/2}\\f_{7/2}\end{matrix}\right.=(n-1)\left\{\begin{matrix}d_{3/2}\\f_{5/2}\\g_{7/2}\\h_{9/2}\end{matrix}\right.\nonumber.\end{equation}
The same behavior is also seen by decreasing the equilibrium inter-nuclear distance by half of the initial. Also, similarly to spin symmetry case, the decrement in $r_e$ result in very small increment in the relativistic energy spectrum. Again, we obtain energy spectrum in the presence of Yukawa tensor term. As it can be seen from the results presented in the table, the degeneracy between the states are removed in this regard. In addition, the system becomes less attractive when the quantum numbers increasing.
\section*{Appendix A. The Nikiforov-Uvarov Method}
\label{APA}
One of the computational tools utilized in solving the Schr\"{o}dinger-like equation including
the centrifugal barrier and/or the spin-orbit coupling term is called the  Nikiforov-Uvarov method (NU). Proposed in order to solve the second-order differential wave equation of the hypergeometric-type:
\begin{equation}
\tag{A1}
\Psi''(z) + \frac{\widetilde{\tau}(z) }{\sigma(z)}\Psi'(z) +
\frac{\widetilde{\sigma}(z) }{\sigma^{2}(z)}\Psi(z) = 0 
\label{A1},
\end{equation}
 where $\sigma(z)$ and $\widetilde{\sigma}(z)$ are polynomials, at most second degree, and $\widetilde{\tau}(z)$  is a polynomial of first degree \cite{HamE101, Ber06, HamE121, HamE122, HamE102, OlE121, OlE122, Ayd09, NiU88, NiE91, YaE10, YaO13, OyE13, Mot09, BeC09, BeE061, EgS05, BeE062, BeH05, BeE05, Ber07, Ber09, IkS07, Ikh09,  MoA10, AkS04, EgE99, EgE00, YaE05, YaE06, TeS09}. 

The prime denotes the differentiation with respect to z. In finding a particular solution to (\ref{A1}), one needs to decompose the wave function $\psi(z)$ as
\begin{equation}
\tag{A2}
\psi(z) = \phi(z)y(z) 
\label{A2},
\end{equation}
yielding the following hypergeometric type equation
\begin{equation}
\tag{A3}
\sigma(z)y''(z) + \tau(z)y'(z) + \Lambda y(z)  = 0 
\label{A3},
\end{equation}
where
\begin{equation}
\tag{A4}
\tau(z) = \widetilde{\tau}(z) + 2 \pi(z)
\label{A4},
\end{equation}
satisfies the condition $\tau'(z)<0$, which will have a negative derivative and $\pi(z)$ is related to the function $\phi(z)$ by
\begin{equation}
\tag{A5}
\pi(z) =  \sigma(z)\frac{d}{dz}[\ln{\phi(z)}] 
\label{A5}.
\end{equation}
The function $\pi$ and the parameter $\lambda$ require for this method are define as follows
\begin{equation}
\tag{A6}
\pi(z) = \left(\frac{\sigma' -  \widetilde{\sigma}  }{2}   \right)
\pm \sqrt{\left(\frac{\sigma' -  \widetilde{\sigma}  }{2}
\right)^{2} -  \widetilde{\sigma}  + k\sigma} 
\label{A6}.
\end{equation}
\begin{equation}
\tag{A7}
\lambda=  k + \pi'(z) 
\label{A7}.
\end{equation}
In other to find the value of k, the expression under the square root must be a square of a polynomial. This gives the polynomial $\pi(z)$  which is dependent on the transformation function z(r). Also the parameter $\lambda$ defined in equation (\ref{A7}) takes the form
\begin{equation}
\tag{A8}
\lambda_{n} =-n\tau'-\left[\frac{n(n - 1)}{2}\sigma''\right],\ \ ~n = 0, 1, 2,...
 \label{A8}
\end{equation}
The polynomial solutions $y_n (z)$ are given by the Rodrigue relation
\begin{equation}
\tag{A9}
y_{n}(z) = \frac{C_{n}}{\rho(z)}\frac{d^{n}}{dz^{n}}[\sigma^{n}(z)\rho(z)]
\label{A9},
\end{equation}
where $C_{n}$ is a normalization constant and the weight function $\rho(z)$ satisfies the following relation
\begin{equation}
\tag{A10}
 \frac{d}{dz}[\sigma(z)\rho(z)] = \tau(z)\rho(z)
\label{A10}.
\end{equation}

\begin{landscape}
\begin{table}[!h]
{\tiny
\caption{The energies ( in units $fm^{-1}$)  in the spin symmetry limit for $M=1fm^{-1}$, $D=15$, $a=0.1$, $C_{s}=0,5$, $r_{e}=0.8, 0.4fm^{-1}$ in atomic units $\hbar=c=\mu=1$}\vspace*{10pt}{
\begin{tabular}{ccccccccccccc}\hline\hline
{}&{}&{}&{}&{}&{}&{}&{}&{}&{}&{}&{}&{}\\[-1.0ex]
$\ell$&$n, \kappa<0$&$nL_{j=\ell+1/2}$&$E_{n, \kappa<0}$ $(A=0)$&$n, \kappa>0$&$nL_{j=\ell-1/2}$&$E_{n, \kappa>0}$ $(A=0)$&$n, \kappa<0$&$nL_{j=\ell+1/2}$&$E_{n, \kappa<0}$ $(A=0.5)$&$n, \kappa>0$&$nL_{j=\ell-1/2}$&$E_{n, \kappa>0}$ $(A=0.5)$\\[2.5ex]\hline\hline
\multicolumn{1}{c}{} & \multicolumn{12}{c}{$C_s=0, r_e=0.8$}\\[1.5ex]\hline
1	&0, -2	&$0p_{3/2}$	&-0.994680673675	&0, 1	&$0p_{1/2}$	&-0.994680673675	&0, -2	&$0p_{3/2}$	&-0.99513873187	&0, 1	&$0p_{1/2}$	&-0.993888275050	\\[1ex]	
2	&0, -3	&$0d_{5/2}$	&-0.987388385545	&0, 2	&$0d_{3/2}$	&-0.987388385545	&0, -3	&$0d_{5/2}$	&-0.988233299165	&0, 2	&$0d_{3/2}$	&-0.986225745520	\\[1ex]	
3	&0, -4	&$0f_{5/2}$	&-0.977137980190	&0, 3	&$0f_{5/2}$	&-0.977137980190	&0, -4	&$0f_{5/2}$	&-0.97836121346	&0, 3	&$0f_{5/2}$	&-0.975605107885	\\[1ex]	
4	&0, -5	&$0g_{9/2}$	&-0.963933376195	&0, 4	&$0g_{7/2}$	&-0.963933376195	&0, -5	&$0g_{9/2}$	&-0.96552993054	&0, 4	&$0g_{7/2}$	&-0.962032400570	\\[1ex]	
1	&1, -2	&$1p_{3/2}$	&-0.987392704065	&1, 1	&$1p_{1/2}$	&-0.987392704065	&1, -2	&$1p_{3/2}$	&-0.98792328953	&1, 1	&$1p_{1/2}$	&-0.986459755860	\\[1ex]	
2	&1, -3	&$1d_{5/2}$	&-0.977723165240	&1, 2	&$1d_{3/2}$	&-0.977723165240	&1, -3	&$1d_{5/2}$	&-0.97866421436	&1, 2	&$1d_{3/2}$	&-0.976418440970	\\[1ex]	
3	&1, -4	&$1f_{7/2}$	&-0.965030658185	&1, 3	&$1f_{5/2}$	&-0.965030658185	&1, -4	&$1f_{7/2}$	&-0.96636085862	&1, 3	&$1f_{5/2}$	&-0.963356764465	\\[1ex]	
4	&1, -5	&$1g_{9/2}$	&-0.949378933000	&1, 4	&$1g_{7/2}$	&-0.949378933000	&1, -5	&$1g_{9/2}$	&-0.95108754925	&1, 4	&$1g_{7/2}$	&-0.947339237365	\\[1ex]	\hline
			\multicolumn{1}{c}{} & \multicolumn{12}{c}{$C_{s}=0$, $r_e=0.4$}\\[1.5ex]\hline											
1	&0, -2	&$0p_{3/2}$	&-0.994575000815	&0, 1	&$0p_{1/2}$	&-0.994575000815	&0, -2	&$0p_{3/2}$	&-0.995056043500	&0, 1	&$0p_{1/2}$	&-0.993745845455	\\[1ex]	
2	&0, -3	&$0d_{5/2}$	&-0.987132159215	&0, 2	&$0d_{3/2}$	&-0.987132159215	&0, -3	&$0d_{5/2}$	&-0.988017930695	&0, 2	&$0d_{3/2}$	&-0.985915834920	\\[1ex]	
3	&0, -4	&$0f_{5/2}$	&-0.976667194300	&0, 3	&$0f_{5/2}$	&-0.976667194300	&0, -4	&$0f_{5/2}$	&-0.977948605220	&0, 3	&$0f_{5/2}$	&-0.975063776210	\\[1ex]	
4	&0, -5	&$0g_{9/2}$	&-0.963184734430	&0, 4	&$0g_{7/2}$	&-0.963184734430	&0, -5	&$0g_{9/2}$	&-0.964856450845	&0, 4	&$0g_{7/2}$	&-0.961196532235	\\[1ex]	
1	&1, -2	&$1p_{3/2}$	&-0.987167626585	&1, 1	&$1p_{1/2}$	&-0.987167626585	&1, -2	&$1p_{3/2}$	&-0.987726126225	&1, 1	&$1p_{1/2}$	&-0.986189096180	\\[1ex]	
2	&1, -3	&$1d_{5/2}$	&-0.977312285560	&1, 2	&$1d_{3/2}$	&-0.977312285560	&1, -3	&$1d_{5/2}$	&-0.978300677660	&1, 2	&$1d_{3/2}$	&-0.975944854180	\\[1ex]	
3	&1, -4	&$1f_{7/2}$	&-0.964368211980	&1, 3	&$1f_{5/2}$	&-0.964368211980	&1, -4	&$1f_{7/2}$	&-0.965763718230	&1, 3	&$1f_{5/2}$	&-0.962614829910	\\[1ex]	
4	&1, -5	&$1g_{9/2}$	&-0.948401747670	&1, 4	&$1g_{7/2}$	&-0.948401747670	&1, -5	&$1g_{9/2}$	&-0.950192903485	&1, 4	&$1g_{7/2}$	&-0.946266072595	\\[1ex]	\hline
			\multicolumn{1}{c}{} & \multicolumn{12}{c}{$C_{s}=5$, $r_e=0.8$}\\[1.5ex]\hline											
1	&0, -2	&$0p_{3/2}$	& 4.00317053373	&0, 1	&$0p_{1/2}$	& 4.00317053373	&0, -2	&$0p_{3/2}$	& 4.0029287387	&0, 1	&$0p_{1/2}$	& 4.00357380957	\\[1ex]	
2	&0, -3	&$0d_{5/2}$	& 4.00804249641	&0, 2	&$0d_{3/2}$	& 4.00804249641	&0, -3	&$0d_{5/2}$	& 4.00756296838	&0, 2	&$0d_{3/2}$	& 4.00868886968	\\[1ex]	
3	&0, -4	&$0f_{5/2}$	& 4.01497524047	&0, 3	&$0f_{5/2}$	& 4.01497524047	&0, -4	&$0f_{5/2}$	& 4.01426159363	&0, 3	&$0f_{5/2}$	& 4.01585663245	\\[1ex]	
4	&0, -5	&$0g_{9/2}$	& 4.02397586139	&0, 4	&$0g_{7/2}$	& 4.02397586139	&0, -5	&$0g_{9/2}$	& 4.02303016318	&0, 4	&$0g_{7/2}$	& 4.02508933373	\\[1ex]	
1	&1, -2	&$1p_{3/2}$	& 4.00540551251	&1, 1	&$1p_{1/2}$	& 4.00540551251	&1, -2	&$1p_{3/2}$	& 4.0052225504	&1, 1	&$1p_{1/2}$	& 4.00570704969	\\[1ex]	
2	&1, -3	&$1d_{5/2}$	& 4.01185375973	&1, 2	&$1d_{3/2}$	& 4.01185375973	&1, -3	&$1d_{5/2}$	& 4.01141639278	&1, 2	&$1d_{3/2}$	& 4.01244650857	\\[1ex]	
3	&1, -4	&$1f_{7/2}$	& 4.02020368469	&1, 3	&$1f_{5/2}$	& 4.02020368469	&1, -4	&$1f_{7/2}$	& 4.01952508655	&1, 3	&$1f_{5/2}$	& 4.02104493729	\\[1ex]	
4	&1, -5	&$1g_{9/2}$	& 4.03057429824	&1, 4	&$1g_{7/2}$	& 4.03057429824	&1, -5	&$1g_{9/2}$	& 4.02965950438	&1, 4	&$1g_{7/2}$	& 4.03165408392	\\[1ex]	\hline
			\multicolumn{1}{c}{} & \multicolumn{12}{c}{$C_{s}=5$, $r_e=0.4$}\\[1.5ex]\hline											
1	&0, -2	&$0p_{3/2}$	& 4.00322054764	&0, 1	&$0p_{1/2}$	& 4.00322054764	&0, -2	&$0p_{3/2}$	& 4.00296888842	&0, 1	&$0p_{1/2}$	& 4.00363806577	\\[1ex]	
2	&0, -3	&$0d_{5/2}$	& 4.00817915484	&0, 2	&$0d_{3/2}$	& 4.00817915484	&0, -3	&$0d_{5/2}$	& 4.00768038478	&0, 2	&$0d_{3/2}$	& 4.00884949119	\\[1ex]	
3	&0, -4	&$0f_{5/2}$	& 4.01523807128	&0, 3	&$0f_{5/2}$	& 4.01523807128	&0, -4	&$0f_{5/2}$	& 4.01449607769	&0, 3	&$0f_{5/2}$	& 4.01615259373	\\[1ex]	
4	&0, -5	&$0g_{9/2}$	& 4.02440431783	&0, 4	&$0g_{7/2}$	& 4.02440431783	&0, -5	&$0g_{9/2}$	& 4.02342131161	&0, 4	&$0g_{7/2}$	& 4.02555988856	\\[1ex]	
1	&1, -2	&$1p_{3/2}$	& 4.00545796030	&1, 1	&$1p_{1/2}$	& 4.00545796030	&1, -2	&$1p_{3/2}$	& 4.00526892236	&1, 1	&$1p_{1/2}$	& 4.00576677707	\\[1ex]	
2	&1, -3	&$1d_{5/2}$	& 4.01201064991	&1, 2	&$1d_{3/2}$	& 4.01201064991	&1, -3	&$1d_{5/2}$	& 4.01155631255	&1, 2	&$1d_{3/2}$	& 4.01262433221	\\[1ex]	
3	&1, -4	&$1f_{7/2}$	& 4.02050032447	&1, 3	&$1f_{5/2}$	& 4.02050032447	&1, -4	&$1f_{7/2}$	& 4.01979510400	&1, 3	&$1f_{5/2}$	& 4.02137262058	\\[1ex]	
4	&1, -5	&$1g_{9/2}$	& 4.03104902087	&1, 4	&$1g_{7/2}$	& 4.03104902087	&1, -5	&$1g_{9/2}$	& 4.03009835517	&1, 4	&$1g_{7/2}$	& 4.03216924169	\\[1ex]	\hline\hline
\end{tabular}\label{tab1} }
\vspace*{-1pt}}
\end{table}
\end{landscape}
\begin{landscape}
\begin{table}[!h]
{\tiny
\caption{The energies (in units $fm^{-1}$)  in the pseudospin symmetry limit for $M=1fm^{-1}$, $D=15$, $a=0.1$, $C_{s}=0,-5$, $r_{e}=0.4, 0.8fm^{-1}$ in atomic units $\hbar=c=\mu=1$}\vspace*{10pt}{
\begin{tabular}{ccccccccccccc}\hline\hline
{}&{}&{}&{}&{}&{}&{}&{}&{}&{}&{}&{}&{}\\[-1.0ex]
$\tilde{\ell}$&$n, \kappa<0$&$\ell, j$&$E_{n, \kappa<0}$ $(A=0)$&$n-1, \kappa>0$&$\ell+2, j+1$&$E_{n-1, \kappa>0}$ $(A=0)$&$n, \kappa<0$&$\ell, j$&$E_{n, \kappa<0}$ $(A=0.5)$&$n-1, \kappa>0$&$\ell+2, j+1$&$E_{n-1, \kappa>0}$ $(A=0.5)$\\[2.5ex]\hline\hline
			\multicolumn{1}{c}{} & \multicolumn{12}{c}{$C_{ps}=0$, $r_e=0.8$}\\[1.5ex]\hline											
1	&1, -1	&$1s_{1/2}$	& 1.00634753849	&0, 2	&$0d_{3/2}$	& 1.00634753849	&1, -1	&$1s_{1/2}$	& 1.00610639381	&0, 2	&$0d_{3/2}$	& 1.00675977512	\\[1ex]	
2	&1, -2	&$1p_{3/2}$	& 1.01327511440	&0, 3	&$0f_{5/2}$	& 1.01327511440	&1, -2	&$1p_{3/2}$	& 1.01276147189	&0, 3	&$0f_{5/2}$	& 1.01397720951	\\[1ex]	
3	&1, -3	&$1d_{5/2}$	& 1.02227632704	&0, 4	&$0g_{7/2}$	& 1.02227632704	&1, -3	&$1d_{5/2}$	& 1.02149997780	&0, 4	&$0g_{7/2}$	& 1.02324344461	\\[1ex]	
4	&1, -4	&$1f_{7/2}$	& 1.03344083138	&0, 5	&$0h_{9/2}$	& 1.03344083138	&1, -4	&$1f_{7/2}$	& 1.03240678672	&0, 5	&$0h_{9/2}$	& 1.03466554547	\\[1ex]	
1	&2, -1	&$2s_{1/2}$	& 1.00919092054	&1, 2	&$1d_{3/2}$	& 1.00919092054	&2, -1	&$2s_{1/2}$	& 1.00903376458	&1, 2	&$1d_{3/2}$	& 1.00944522248	\\[1ex]	
2	&2, -2	&$2p_{3/2}$	& 1.01837933133	&1, 3	&$1f_{5/2}$	& 1.01837933133	&2, -2	&$2p_{3/2}$	& 1.01790936762	&1, 3	&$1f_{5/2}$	& 1.01902421126	\\[1ex]	
3	&2, -3	&$2d_{5/2}$	& 1.02909646038	&1, 4	&$1g_{7/2}$	& 1.02909646038	&2, -3	&$2d_{5/2}$	& 1.02835323755	&1, 4	&$1g_{7/2}$	& 1.0300254518	\\[1ex]	
4	&2, -4	&$2f_{7/2}$	& 1.04185296565	&1, 5	&$1h_{9/2}$	& 1.04185296565	&2, -4	&$2f_{7/2}$	& 1.04084597013	&1, 5	&$1h_{9/2}$	& 1.04304855818	\\[1ex]	\hline
			\multicolumn{1}{c}{} & \multicolumn{12}{c}{$C_{ps}=0$, $r_e=0.4$}\\[1.5ex]\hline											
1	&1, -1	&$1s_{1/2}$	& 1.00642272478	&0, 2	&$0d_{3/2}$	& 1.00642272478	&1, -1	&$1s_{1/2}$	& 1.00617161853	&0, 2	&$0d_{3/2}$	& 1.00684974812	\\[1ex]	
2	&1, -2	&$1p_{3/2}$	& 1.01346626112	&0, 3	&$0f_{5/2}$	& 1.01346626112	&1, -2	&$1p_{3/2}$	& 1.01293096354	&0, 3	&$0f_{5/2}$	& 1.01419592001	\\[1ex]	
3	&1, -3	&$1d_{5/2}$	& 1.02262388110	&0, 4	&$0g_{7/2}$	& 1.02262388110	&1, -3	&$1d_{5/2}$	& 1.02181504291	&0, 4	&$0g_{7/2}$	& 1.02362950439	\\[1ex]	
4	&1, -4	&$1f_{7/2}$	& 1.03398700259	&0, 5	&$0h_{9/2}$	& 1.03398700259	&1, -4	&$1f_{7/2}$	& 1.03291001560	&0, 5	&$0h_{9/2}$	& 1.03526065500	\\[1ex]	
1	&2, -1	&$2s_{1/2}$	& 1.00925346428	&1, 2	&$1d_{3/2}$	& 1.00925346428	&2, -1	&$2s_{1/2}$	& 1.00909303459	&1, 2	&$1d_{3/2}$	& 1.00950917764	\\[1ex]	
2	&2, -2	&$2p_{3/2}$	& 1.01859755340	&1, 3	&$1f_{5/2}$	& 1.01859755340	&2, -2	&$2p_{3/2}$	& 1.01810852148	&1, 3	&$1f_{5/2}$	& 1.01926660631	\\[1ex]	
3	&2, -3	&$2d_{5/2}$	& 1.02949160803	&1, 4	&$1g_{7/2}$	& 1.02949160803	&2, -3	&$2d_{5/2}$	& 1.02871759605	&1, 4	&$1g_{7/2}$	& 1.03045714342	\\[1ex]	
4	&2, -4	&$2f_{7/2}$	& 1.04246400301	&1, 5	&$1h_{9/2}$	& 1.04246400301	&2, -4	&$2f_{7/2}$	& 1.04141529809	&1, 5	&$1h_{9/2}$	& 1.04370719456	\\[1ex]	\hline
			\multicolumn{1}{c}{} & \multicolumn{12}{c}{$C_{ps}=-5$, $r_e=0.8$}\\[1.5ex]\hline											
1	&1, -1	&$1s_{1/2}$	&-3.98474235212	&0, 2	&$0d_{3/2}$	&-3.98474235212	&1, -1	&$1s_{1/2}$	&-3.98536953322	&0, 2	&$0d_{3/2}$	&-3.98363892498	\\[1ex]	
2	&1, -2	&$1p_{3/2}$	&-3.97408332396	&0, 3	&$0f_{5/2}$	&-3.97408332396	&1, -2	&$1p_{3/2}$	&-3.97517771342	&0, 3	&$0f_{5/2}$	&-3.97256443759	\\[1ex]	
3	&1, -3	&$1d_{5/2}$	&-3.96005901271	&0, 4	&$0g_{7/2}$	&-3.96005901271	&1, -3	&$1d_{5/2}$	&-3.96158951648	&0, 4	&$0g_{7/2}$	&-3.95813105164	\\[1ex]	
4	&1, -4	&$1f_{7/2}$	&-3.94279155689	&0, 5	&$0h_{9/2}$	&-3.94279155689	&1, -4	&$1f_{7/2}$	&-3.94474354082	&0, 5	&$0h_{9/2}$	&-3.94045911429	\\[1ex]	
1	&2, -1	&$2s_{1/2}$	&-3.96931123016	&1, 2	&$1d_{3/2}$	&-3.96931123016	&2, -1	&$2s_{1/2}$	&-3.96999215207	&1, 2	&$1d_{3/2}$	&-3.96810323038	\\[1ex]	
2	&2, -2	&$2p_{3/2}$	&-3.95679768446	&1, 3	&$1f_{5/2}$	&-3.95679768446	&2, -2	&$2p_{3/2}$	&-3.95798950402	&1, 3	&$1f_{5/2}$	&-3.95513593304	\\[1ex]	
3	&2, -3	&$2d_{5/2}$	&-3.94046656480	&1, 4	&$1g_{7/2}$	&-3.94046656480	&2, -3	&$2d_{5/2}$	&-3.9421162245	&1, 4	&$1g_{7/2}$	&-3.93838260791	\\[1ex]	
4	&2, -4	&$2f_{7/2}$	&-3.92071895908	&1, 5	&$1h_{9/2}$	&-3.92071895908	&2, -4	&$2f_{7/2}$	&-3.92280064576	&1, 5	&$1h_{9/2}$	&-3.91822686782	\\[1ex]	\hline
			\multicolumn{1}{c}{} & \multicolumn{12}{c}{$C_{ps}=-5$, $r_e=0.4$}\\[1.5ex]\hline											
1	&1, -1	&$1s_{1/2}$	&-3.98445312686	&0, 2	&$0d_{3/2}$	&-3.98445312686	&1, -1	&$1s_{1/2}$	&-3.98511403062	&0, 2	&$0d_{3/2}$	&-3.98329452394	\\[1ex]	
2	&1, -2	&$1p_{3/2}$	&-3.97358257158	&0, 3	&$0f_{5/2}$	&-3.97358257158	&1, -2	&$1p_{3/2}$	&-3.97473339040	&0, 3	&$0f_{5/2}$	&-3.97198877105	\\[1ex]	
3	&1, -3	&$1d_{5/2}$	&-3.95927228873	&0, 4	&$0g_{7/2}$	&-3.95927228873	&1, -3	&$1d_{5/2}$	&-3.96087984589	&0, 4	&$0g_{7/2}$	&-3.95725033962	\\[1ex]	
4	&1, -4	&$1f_{7/2}$	&-3.94164843833	&0, 5	&$0h_{9/2}$	&-3.94164843833	&1, -4	&$1f_{7/2}$	&-3.94369711585	&0, 5	&$0h_{9/2}$	&-3.93920334486	\\[1ex]	
1	&2, -1	&$2s_{1/2}$	&-3.9687414894	&1, 2	&$1d_{3/2}$	&-3.9687414894	&2, -1	&$2s_{1/2}$	&-3.96945941660	&1, 2	&$1d_{3/2}$	&-3.96747211076	\\[1ex]	
2	&2, -2	&$2p_{3/2}$	&-3.95599216066	&1, 3	&$1f_{5/2}$	&-3.95599216066	&2, -2	&$2p_{3/2}$	&-3.95724655890	&1, 3	&$1f_{5/2}$	&-3.95424678741	\\[1ex]	
3	&2, -3	&$2d_{5/2}$	&-3.93934209634	&1, 4	&$1g_{7/2}$	&-3.93934209634	&2, -3	&$2d_{5/2}$	&-3.94107637432	&1, 4	&$1g_{7/2}$	&-3.93715457112	\\[1ex]	
4	&2, -4	&$2f_{7/2}$	&-3.91920194081	&1, 5	&$1h_{9/2}$	&-3.91920194081	&2, -4	&$2f_{7/2}$	&-3.92138852243	&1, 5	&$1h_{9/2}$	&-3.91658741001	\\[1ex]	\hline\hline
\end{tabular}\label{tab2} }
\vspace*{-1pt}}
\end{table}
\end{landscape}

\end{document}